\documentclass[times,trackchanges]{aastex631} % twocolumn linenumbers,

\submitjournal{ApJ}

\usepackage{savesym}

\restoresymbol{SIX}{tablenum}
\usepackage{booktabs}
\usepackage{multirow}
\usepackage{physics}
\usepackage{textcomp}
\usepackage{hyperref}
\usepackage[capitalize]{cleveref}
\usepackage{amssymb}
\usepackage{xcolor}
\usepackage{graphicx}  % for \rotatebox
\usepackage{longtable} % 
\usepackage{makecell} % 
\usepackage{tabularx}

\makeatletter
\usepackage{etoolbox}
\patchcmd\H@refstepcounter{\protected@edef}{\protected@xdef}{}{}
\makeatother

\graphicspath{{./}{assets/}}

\begin{document}

% ---- Front matter
\title{Surface Waves at Switchback Boundaries in the Young Solar Wind from Parker Solar Probe Observations}
\shorttitle{Surface Waves at Switchback Boundaries}
\shortauthors{Choi et al.}

% ---- Authors
\author[0000-0003-2054-6011]{Kyung-Eun Choi}
\affiliation{Space Sciences Laboratory, University of California Berkeley: Berkeley, CA, USA}
\author[0000-0001-6427-1596]{Oleksiy V. Agapitov}
\affiliation{Space Sciences Laboratory, University of California Berkeley: Berkeley, CA, USA}
\affiliation{Astronomy and Space Physics Department, National Taras Shevchenko University of Kyiv, Kyiv, Ukraine}
\author[0000-0001-6767-0672]{Nina Bizien}
\affiliation{LPC2E, OSUC, Univ Orleans, CNRS, CNES, F-45071 Orleans, France}
\author[0000-0002-4401-0943]{Thierry {Dudok de Wit}}
\affiliation{LPC2E, OSUC, Univ Orleans, CNRS, CNES, F-45071 Orleans, France}
\affiliation{International Space Science Institute, ISSI, Bern, Switzerland}
\author[0000-0001-6016-7548]{Lucas Colomban}
\affiliation{Space Sciences Laboratory, University of California Berkeley: Berkeley, CA, USA}

%Forrest? 
% ..................

\correspondingauthor{Kyung-Eun Choi}
\email{kechoi@berkeley.edu}

% ---- Abstract
\begin{abstract}
Switchbacks (SBs) are localized magnetic field deflections in the solar wind, marked by abrupt changes in the magnetic field direction relative to the ambient solar wind. Observations onboard Parker Solar Probe (PSP) at heliocentric distances below 50 Solar Radii \(R_S\) showed that within SBs, perturbations in the magnetic field ($\Delta \vec{B}$) and the bulk solar wind velocity ($\Delta \vec{V}$) align, i.e., $\Delta \vec{B} \sim \Delta \vec{V}$, producing enhanced radial velocity spikes. %Surface waves have been suggested to play a role in the enhanced wave activity observed at SB boundaries.
In this study, we examine the characteristics of SB boundaries, with particular attention to the role of boundary shear flow instabilities (Kelvin-Helmholtz instability - KHI) for surface wave phenomena based on the \textit{in situ} magnetic field, plasma speed, and plasma density measurements from PSP.
%The stability of these boundary layers is evaluated in the context of the Kelvin-Helmholtz (KH) instability driven by velocity shear across the boundary.%to reformulate
%We report PSP observations of surface waves occurring at SB boundaries. 
%While previous studies have hypothesized the development of KH instability at such shear interfaces, direct in-situ evidence has been lacking. 
The results indicate that SB boundaries can be unstable for generating KHI-driven surface waves, suggesting that the wave activity observed at SB boundaries is caused by shear flow instabilities. In addition, the continued development of KHI may lead to boundary erosion, contributing to the radial evolution of SBs via structural weakening or broadening.
%However, if $\Delta \vec{B} =\pm \alpha  \Delta \vec{V}$, then for $\alpha < 2$ (usually, for SBs,  $0.4<\alpha < 1.1$), the boundary is stable, and the observed bursts of wave activity should be the remnant of the surface instability. 
However, when $\Delta \vec{B}$ and $\Delta \vec{V}$ are closely aligned, the boundary remains stable unless the velocity shear significantly exceeds the magnetic shear. Since the observed velocity shear typically ranges from 40\% to 90\% of magnetic shear, the instability condition is generally not satisfied.
Thus, the configuration leading to the instability arises from deviations from precise alignment of $\Delta \vec{B}$ and  $\Delta \vec{V}$ in the young solar wind, and the release of the KHI presumably leads to the formation of the $\Delta \vec{B}$ and  $\Delta \vec{V}$ alignment observed at SB boundaries located at 35-55 \(R_S\).  
\end{abstract}


% ---- Main text
\section{Introduction}\label{sec:intro}    % [-]

%-SBs, definition, observed by the PSP and others, origins... [-]
Among the various features identified by Parker Solar Probe (PSP) in the near-Sun solar wind, magnetic field switchbacks (SBs) have received particular attention. 
SBs are characterized by localized, large-angle deflections of the magnetic field direction relative to the ambient solar wind.
A threshold of 45$^\circ$ is usually applied to identify SBs. 
These deflections generally occur without magnetic field magnitude changes, although depletion or enhancement less than $0.1 B_0$ are observed in some events. Their distinctive signatures \citep{bale+2019,kasper+2019, agapitov+2023, krasnoselskikh+2020} and potential implications for solar wind dynamics suggest that they may contribute to magnetic and plasma energy release and local solar wind heating.
Within SBs, perturbations in the magnetic field ($\Delta \vec{B}$) and the bulk solar wind velocity ($\Delta \vec{V}$) are often aligned, so that $\Delta \vec{B} \sim \Delta \vec{V}$ if the background magnetic field direction is toward the Sun (or $\Delta \vec{B} \sim -\Delta \vec{V}$ if the background magnetic field direction is directed from the Sun), leading to localized enhancements in the radial component of the bulk velocity \citep[][]{kasper+2019, bale+2019, mozer+2020, krasnoselskikh+2020}.
Although sudden deviations of magnetic field direction from the Parker spiral have been reported in earlier solar wind observations across the heliosphere \citep[e.g.,][]{kahler_lin_1994, neugebauer+1995, kahler+1996, crooker+2004, owens+2013, huang+2017, horbury+2018, woolley+2021, Zhao+2025}, the high occurrence of these large-angle deflections in the young solar wind, as revealed by recent PSP observations, has brought renewed attention to their nature and origin \citep[e.g.,][]{bale+2019, kasper+2019, horbury+2020, dudok_de_wit+2020, mozer+2020}. 
The origin of SBs remains under debate \citep{raouafi+2023}. Proposed mechanisms include solar-surface processes such as interchange reconnection and jet-like eruptions from coronal bright points \citep[e.g.,][]{horbury+2020, Zank+2020, Drake+2021, Neugebauer_Sterling_2021, telloni+2022, Fargette+2022, Lee+2024, Bizien+2025}, as well as in situ formation via turbulence, velocity shear, or solar wind expansion \citep[e.g.,][]{Ruffolo+2020, Squire+2020, schwadron_and_McComas_2021, toth+2023, Jagarlamudi+2023}. 

%-SB structure - as a tube and frozen-in (e.g. Krasnoselskikh et al.2020)% [-]
SBs are frequently observed to exhibit Alfvénic properties, characterized by strong alignment between magnetic field and plasma velocity perturbations \citep{kasper+2019, bale+2019, larosa+2021, krasnoselskikh+2020}. 
While such Alfvénic signatures are common, a subset of SBs also displays coherent internal magnetic structures that persist over several minutes. 
This suggests that these SBs may retain a spatially coherent magnetic geometry, which may correspond to a localized structure rather than a transient wave packet. 
\cite{horbury+2020} suggested that SBs preserve their internal field orientation over several minutes, consistent with the propagation of flux-tube-like entities. \cite{krasnoselskikh+2020} interpreted the frozen-in character of SBs as magnetic flux tube-like structures, bounded by sharp current layers where the magnetic field direction inside differs significantly from that of the surrounding solar wind.

\begin{figure}
    \centering
    \includegraphics[width=\textwidth]{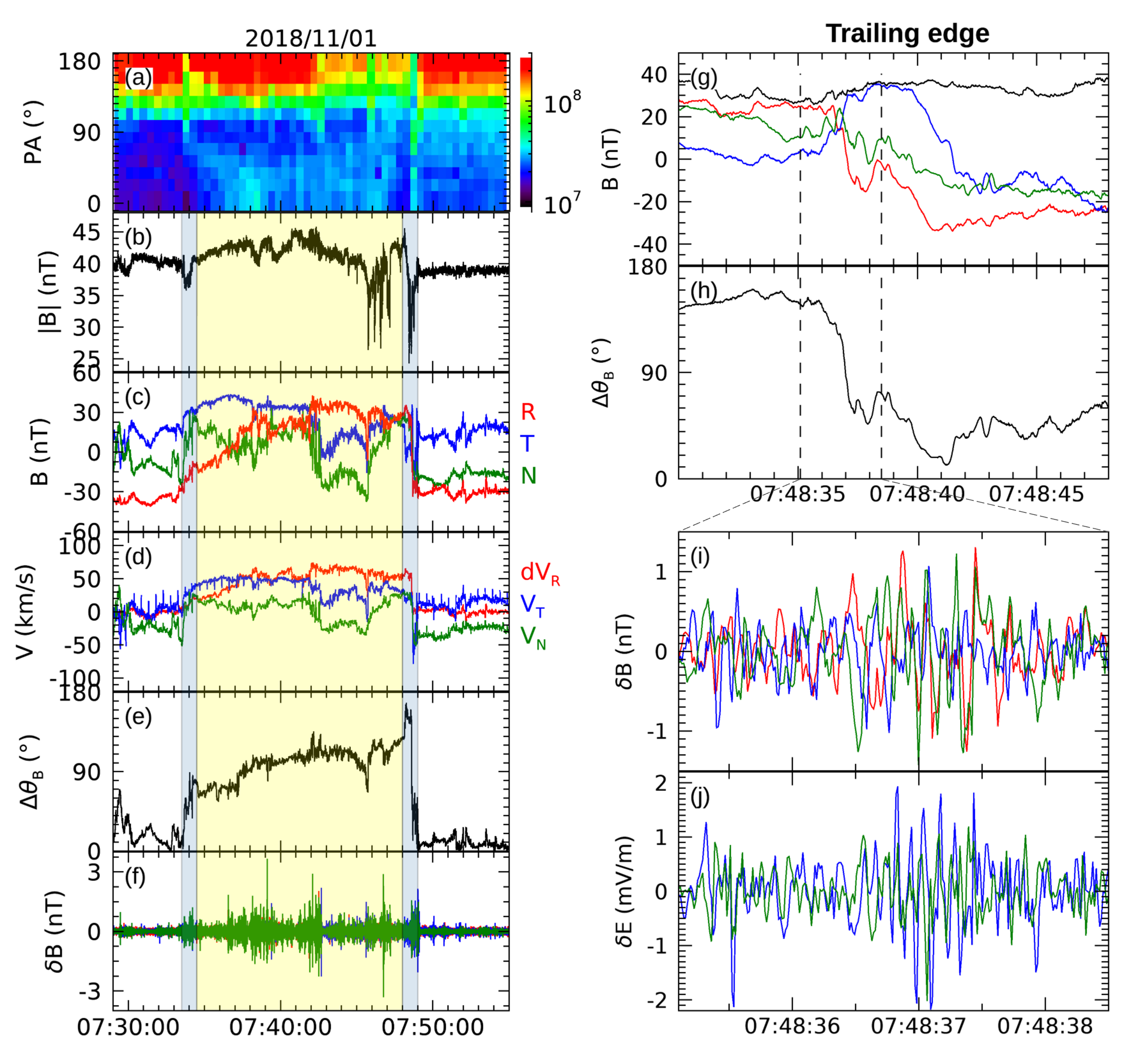}   
    \caption{An example of an SB recorded by PSP on November 1, 2018. Panels (a)–(f) show the overall structure of the SB, including its boundaries (the blue-shaded regions) and the main spike (the yellow-shaded region); (a) - pitch angle distribution of suprathermal electrons (314.5 eV); (b) - magnetic field magnitude from MAG data; (c) - magnetic field components in the $RTN$ coordinate system; (d) - velocity components in $RTN$, with $dV_R = V_R - V_{R0}$, where $V_{R0}$ represents the ambient solar wind bulk velocity; (e) - angular difference of the magnetic field relative to the background solar wind; (f) - magnetic field perturbations from SCM data. Panels (g) and (h) provide a zoomed-in view of magnetic field components and angular differences at the trailing boundary near 07:48 UT, respectively. Panels (i) and (j) display wave activity within the time interval between the two dashed vertical lines in panels (g) and (h): (i) - magnetic field wave activity; (j) - electric field wave activity.}
   \label{fig:overview}
\end{figure}

%- Boundary of SBs - (Kasper + 2019, Mozer + 2020, Farrell et al. 2020, Larosa + 2021, Bizien + 2023) [-]
Recent studies have noted structural features at SB boundaries \citep[e.g.,][]{kasper+2019, mozer+2020, farrell+2020, larosa+2021, Martinović+2021,rasca+2021,rasca+2022,bizien+2023}. In addition to these structural properties, several distinct physical signatures, including magnetic field dropouts, discontinuity-like structures, and enhanced wave activity, have also been reported \citep{farrell+2020, farrell+2021}.
%1. Magnetic field dropouts. 
The first aspect is the occurrence of magnetic field dropouts (B-dropouts; localized B decrease) at SB boundaries \citep[e.g.,][]{farrell+2020, krasnoselskikh+2020,agapitov+2020}. \cite{farrell+2020} reported this feature using superposed epoch analysis, which revealed systematic depletions in the magnetic field magnitude, a feature that is also evident in the example we shall address below (see \Cref{fig:overview}(b) for reference). These B-dropouts are typically observed near sharp magnetic transitions, suggesting that they may be related to boundary-layer structures. The presence of such localized field gradients may also imply the existence of thin current sheets at the boundaries. In a related context, \cite{huang+2023} proposed that thin current sheets formed by magnetic braiding may help stabilize the SB structure, potentially contributing to the observed magnetic features at the boundary.
%2. Most of the observed boundaries in switchback events can be classified as tangential discontinuities. (Bizien + 2023)
A second characteristic is that many boundaries exhibit features consistent with tangential discontinuities \citep[TDs;][]{hudson1970}. This perspective has been adopted in several recent studies \citep[e.g.,][]{larosa+2021, bizien+2023}, which have analyzed SB boundaries in terms of discontinuity properties. \cite{bizien+2023} showed that the majority of boundaries analyzed display sharp magnetic rotations with negligible normal components, supporting the TD interpretation. In contrast, \cite{larosa+2021} reported that some events may resemble rotational discontinuities (RDs), indicating variability across different conditions.
%3. Enhanced wave activities are observed at the boundary of SBs.
A third feature of SB boundaries is the presence of enhanced wave activity. In particular, whistler-mode waves have been reported to occur near these boundaries and are often associated with localized magnetic field depletion \citep[][]{agapitov+2020, jagarlamudi+2021whistler, froment+2023whistler, karbashewski=2023whistler, colomban+2024, choi+2024whistler}. In addition to these high-frequency waves, \cite{krasnoselskikh+2020} identified lower-frequency fluctuations, ranging from approximately 0.3 to 10 Hz in the spacecraft frame, which have been interpreted as surface waves. 

Surface waves are known to occur on solar wind TDs, and their properties and propagation conditions were described by \cite{Hollweg1982}. Given that many SBs' boundaries exhibit TD-like features \citep{bizien+2023}, similar surface wave phenomena may arise in association with SBs. Surface waves, inferred from low-frequency fluctuations near SB boundaries \citep[e.g.,][]{krasnoselskikh+2020}, have been proposed as a possible contributor to the enhanced wave activity occasionally observed at these boundaries. According to \cite{mozer+2020}, surface waves observed near the boundaries of SBs may arise from the Kelvin–Helmholtz (KH) instability, as driven by velocity shear between the faster ion flow inside the SBs and the slower ambient solar wind. However, the growth timescales of the KH instability are known to be too short to allow direct in-situ detection, thereby leaving its observational confirmation unresolved. 

In this study, we investigate the presence of surface waves at SB boundaries and explore whether they can be attributed to the KH instability. We address this by examining whether the observed fluctuations exhibit surface wave characteristics and by evaluating the instability condition through the KH threshold. The structure of this paper is as follows. Section \ref{sec:sec2} presents the observational results, including surface wave signatures at SB boundaries and their basic properties. Section \ref{sec:dis_conc} provides a discussion of the potential implications of these observations, introduces additional case studies, and examines whether the observed fluctuations can be explained by the KH instability. We conclude with a summary of the key findings and remarks on possible directions for future work.

\section{Observations and Analysis}\label{sec:sec2}  % [-]

This study employs magnetic field and plasma data from the Parker Solar Probe (PSP). Magnetic and electric field data are obtained from the PSP Fluxgate Magnetometer (MAG) and the Electric Field Instrument \citep[EFI;][]{mozer+2020} of the FIELDS experiment \citep{bale+2016}. The plasma data are provided by the Solar Wind Electrons Alphas and Protons suite \citep[SWEAP;][]{kasper+2016}, which includes the Solar Probe Cup \citep[SPC;][]{case+2020} and Solar Probe Analyzers \citep[SPAN-i and SPAN-e;][]{livi+2022,whittlesey+2020}. 
% and Search Coil Magnetometer (SCM, ref) 

All measurements are expressed in the RTN coordinate system (Radial–Tangential–Normal), where the R-axis points radially away from the Sun, the T-axis is oriented in the direction of spacecraft motion, and the N-axis completes the right-handed set. For the purpose of describing the directional properties of field and flow vectors, we also employ a spherical coordinate system defined with respect to the RTN frame. In this system, $\theta$ denotes the polar angle measured from the radial (R) direction, and $\phi$ represents the azimuthal angle measured in the T–N plane. This spherical coordinate system is primarily used in the analysis presented in \Cref{fig:KHtest}.

\subsection{Switchback Boundaries and Their Structure} % [-]

\Cref{fig:overview} presents a typical SB that illustrates the magnetic field and velocity perturbation structures near the boundaries and the associated wave activities, as recorded onboard PSP on November 1, 2018. During the interval (07:34–07:49 UT), the strahl maintains an anti-parallel direction (i.e., the electron beam appears at 180$^\circ$ in the pitch-angle distribution), while the magnetic field rotates significantly. This SB exhibits an average angular deflection of approximately 90$^\circ$ (with respect to the ambient solar wind) and peaks at 160$^\circ$ at the trailing edge. Notably, $B_R$ reverses its sign, whereas the field magnitude $|B|$ remains relatively constant throughout the interval. The velocity profile shows a similar rotational behavior. It is noteworthy that the field magnitude $|B|$ shows clear depletion at both edges, as indicated by the blue shaded regions in the left column of \Cref{fig:overview}, which indicates the presence of current sheets at the boundaries \cite{agapitov+2022, krasnoselskikh+2020, larosa+2021}. \Cref{fig:overview}(f) presents the magnetic field perturbations ($\delta B$) that indicate wave bursts near both the leading and trailing boundaries, as well as intermittently within the SB interval. 
The right column of \Cref{fig:overview} provides a zoom-in view of the trailing edge, presenting the magnetic field (Figure \ref{fig:overview}(g)), magnetic field deflection angle ($\Delta\theta_B$, Figure\ref{fig:overview}(h)) with respect to the ambient solar wind, and the associated with the boundary wave activity observed in magnetic ($\delta B$) and electric field ($\delta E$) measurements (Figures \ref{fig:overview}(i) and \ref{fig:overview}(j)) - the subject of our study.

\subsection{Surface Waves at the Boundary Layer: Observational Signatures} % [-]

\begin{figure}
    \centering
    \includegraphics[width=\textwidth]{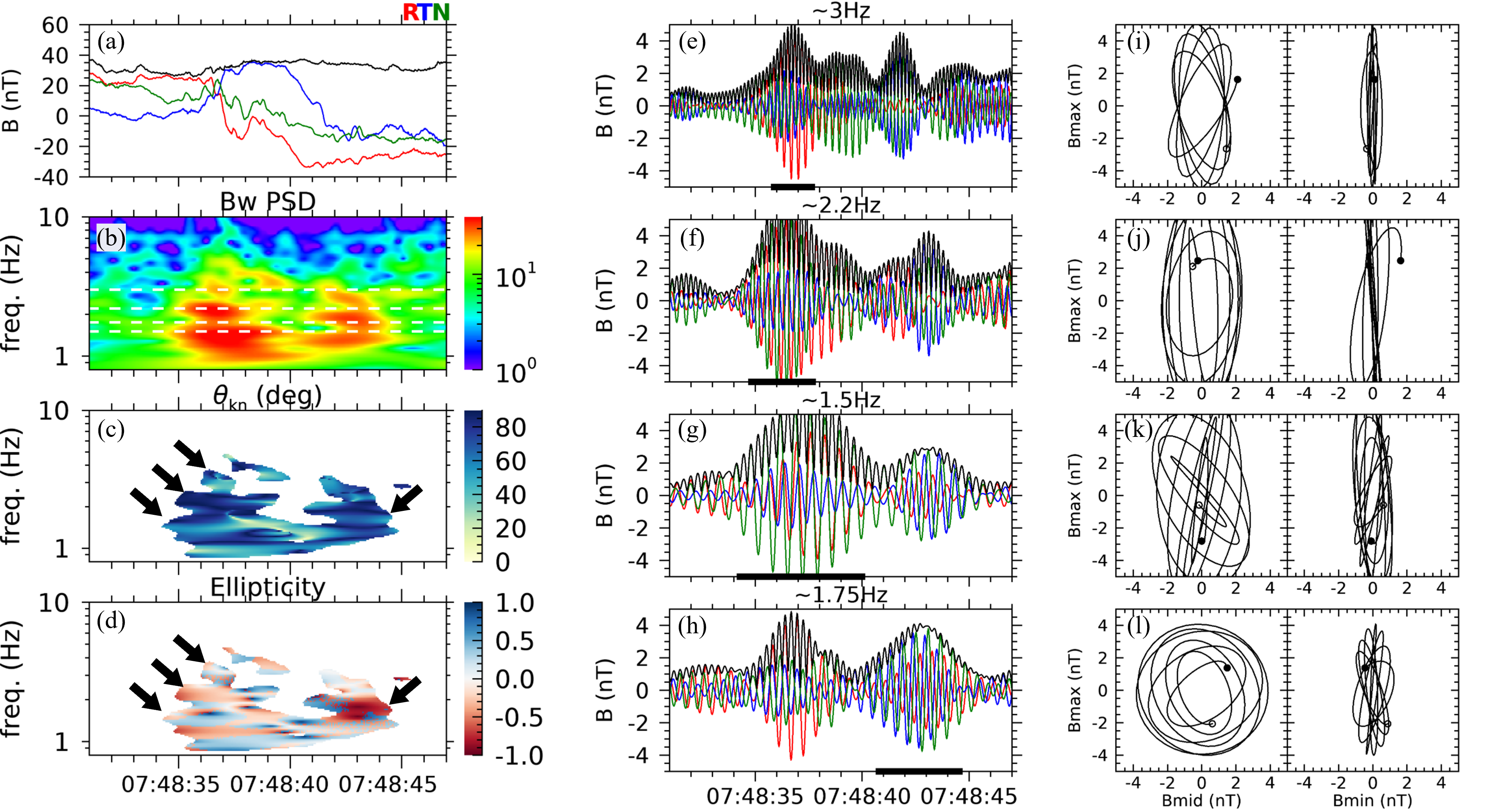}
    \caption{Wave activity at the trailing boundary of the SB in \Cref{fig:overview}.
    The first column presents the magnetic field variation (panel (a)) and wave properties. The power spectral density (PSD), wave propagation direction ($\vb*{k}$) relative to the normal vector ($\vb*{n}$) of the layer, and signed ellipticity are shown in panels (b)–(d), respectively.  
    The black arrows in panels (c) and (d) mark intervals of enhanced wave activity where the propagation angle $\theta_{kn}$ exceeds 80$^\circ$, consistent with the blue features in panel (c), and matching the frequency bands used for filtering in panels (e)–(h).
    The second column displays the magnetic field waveforms in the RTN frame, band-pass filtered around the frequencies indicated by the white dashed horizontal lines in panel (b): 3, 2.2, 1.5, and 1.75~Hz, respectively.
    The last column shows hodographs of the magnetic field during the intervals indicated by the thick horizontal bars in panels (e)-(h), revealing polarization characteristics — linear, elliptical, or circular — of the observed waves. 
    } 
   \label{fig:wave}
\end{figure}

\Cref{fig:wave} presents detailed signatures of wave activity observed at the trailing edge of the SB. \Cref{fig:wave}(a) shows a sharp magnetic field variation across the boundary layer. 
The boundary normal vector ($\vb*{n}$) is determined using minimum variance analysis (MVA), yielding a direction of  $(n_R, n_T, n_N) = (0.48,0.21, -0.85)$. This normal vector serves as a reference for evaluating the propagation direction of the wave, as examined later in \Cref{fig:wave}(c). \Cref{fig:wave}(b) shows the power spectral density, where enhanced wave activity is evident near 1.5, 1.75, 2.2, and 3 Hz (indicated by arrows). These frequencies correspond to the blue-colored features in \Cref{fig:wave}(c), where the angle between the wave vector $\vb*{k}$ and boundary normal $\vb*{n}$ (denoted as $\theta_{kn}$) exceeds 80$^\circ$, indicating propagation nearly parallel to the boundary surface. 

\Cref{fig:wave}(d) shows the signed ellipticity of the waves \citep{santolik+2003}.
At 07:48:36 UT, three enhanced wave structures near 1.5, 2.2, and 3~Hz exhibit ellipticity values less than 0.2, consistent with linear polarization. Together with the nearly perpendicular $\vb*{k}$ directions shown in \Cref{fig:wave}(c), these waves exhibit key characteristics of surface waves, namely (i) surface-aligned propagation and (ii) linear polarization.
These features are further supported by bandpass-filtered waveforms shown in Figures \ref{fig:wave}(e–g), and the corresponding hodographs in Figures \ref{fig:wave}(i–k), which confirm polarization states ranging from linear to weakly elliptical.

In contrast, a different wave signature is observed at 07:48:43 UT, shortly after the surface waves. This wave, enhanced near 1.75~Hz, is left-hand circularly polarized with an ellipticity close to $-1$, as shown in \Cref{fig:wave}(d). Interestingly, it also exhibits a nearly perpendicular propagation direction, similar to the surface waves. Although surface waves can occasionally be circularly polarized \citep{Hollweg1982}, the distinct polarization and temporal separation suggest that this wave may represent a different mode. This point will be discussed in more detail in Section \ref{sec:dis_conc}. The corresponding waveform and hodograph are presented in Figures \ref{fig:wave}(h) and \ref{fig:wave}(l), respectively.

\subsection{Examination of Kelvin-Helmholtz Instability Criteria} % [-]

To evaluate whether the characteristics of the observed surface wave are consistent with a KH-driven origin, we examine the instability criteria under the observed boundary conditions based on the classical MHD formulation \citep{miura1984}. The instability criterion between two different regimes is given by:
\begin{equation}
    \rho_1\rho_2[(\vb*{v_1}  - \vb*{v_2} )\cdot\vb*{k}]^2>(\rho_1+\rho_2)[(\vb*{B_1}\cdot\vb*{k})^2 + (\vb*{B_2}\cdot\vb*{k})^2]/\mu_0,
\end{equation}
where $\vb*{v}_{1,2}$ and $\vb*{B}_{1,2}$ denote the plasma velocity and magnetic field on each side of the boundary, $\rho_{1,2}$ are the corresponding mass densities, and $\vb*{k}$ is the wave vector. 
For visualization and comparative analysis, we define a scalar KHI threshold as follows:
\begin{equation}
  \mathrm{KHI\,threshold} \equiv \rho_1\rho_2[(\vb*{v_1}  - \vb*{v_2} )\cdot\vb*{k}]^2 - (\rho_1+\rho_2)[(\vb*{B_1}\cdot\vb*{k})^2 + (\vb*{B_2}\cdot\vb*{k})^2]/\mu_0, 
\end{equation}
such that a positive value implies instability for the given wave vector direction.
We survey all possible directions of $\vb*{k}$ to determine whether they lie within stable or unstable regions under the given shear conditions. The calculation is performed using measured solar wind parameters inside and outside the SB boundary, which is treated as a shear layer. 

\begin{figure}
    \centering
    \includegraphics[]{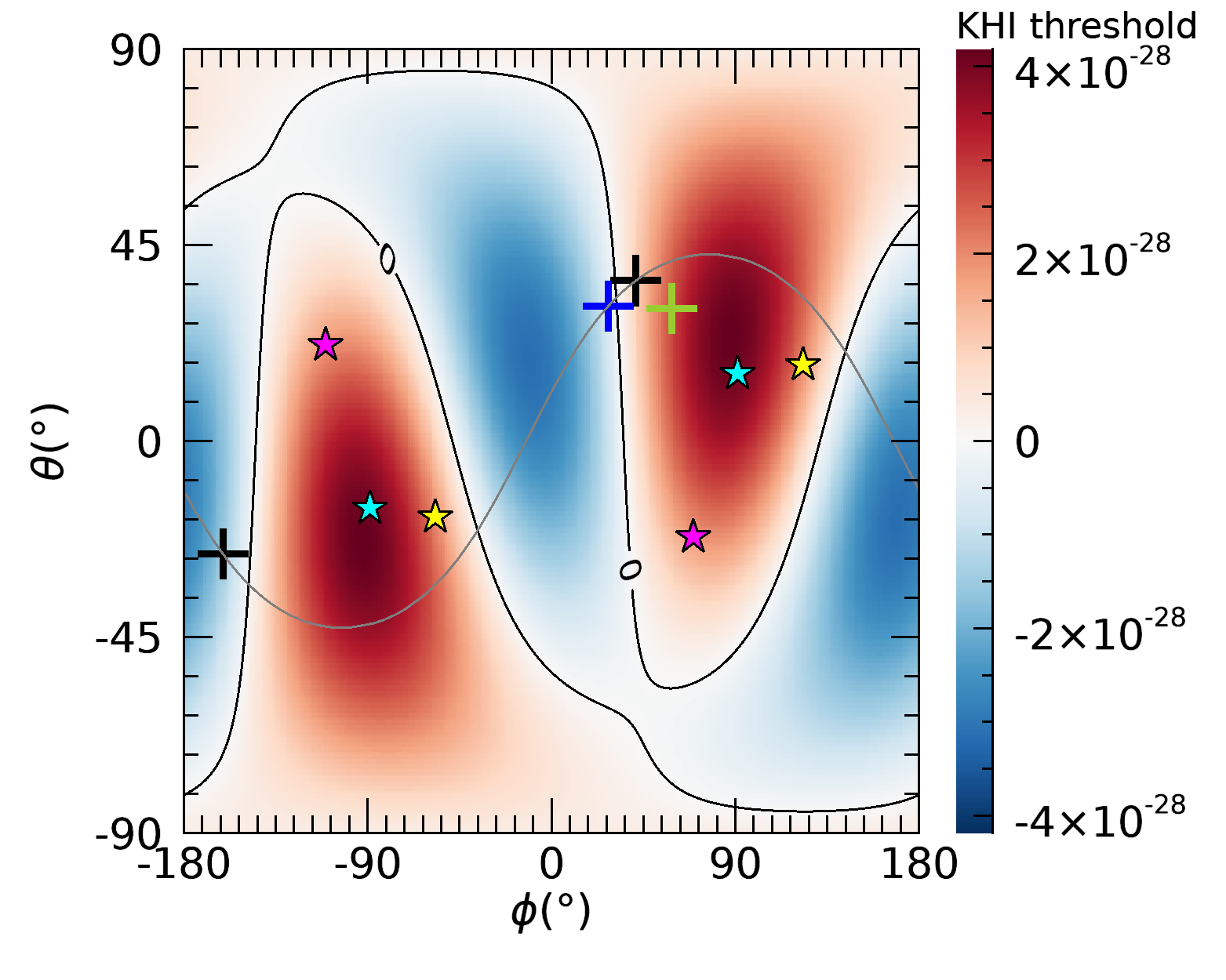}
    \caption{Kelvin-Helmholtz instability growth rate and directional distribution of plasma parameters. The '+' symbols mark the magnetic fields ($\mathbf{B}_1$ and $\mathbf{B}_2$, black), their difference ($\Delta \vec{B} = \mathbf{B}_1 - \mathbf{B}_2$, blue), and the velocity difference ($\Delta \vec{V} = \mathbf{V}_1 - \mathbf{V}_2$, green). The gray curve outlines the plane defined by $\mathbf{B}_1$, $\mathbf{B}_2$, and $\Delta \vec{B}$. The background color map shows the KHI threshold (see Equation~(2)) as a function of wave vector direction in spherical coordinates ($\phi$, $\theta$). Asterisks in three colors indicate the wave vector directions of surface waves in \Cref{fig:wave} at 1.5~Hz (yellow), 2.2~Hz (cyan), and 3~Hz (magenta), each showing both forward and backward propagation.}
   \label{fig:KHtest}
\end{figure}

\Cref{fig:KHtest} summarizes the directional dependence of the KHI threshold at the trailing edge of the SB event shown in \Cref{fig:overview}, with wave vector directions and measured plasma parameters superposed for comparison. If all directions of $\vb*{k}$ result in stable conditions (i.e. negative values in the KHI threshold calculation, shown in blue), then the linearly polarized surface waves in that layer are not necessarily evidence of KH instability. Instead, they may reflect a remnant of a previously unstable, now stabilized boundary. Thus, the possibility of a KH-driven origin is not excluded.
However, in the present case, some directions of $\vb*{k}$ correspond to positive threshold values, indicating that instability may occur (shown in red). 
The wave vectors of the three surface wave events analyzed in \Cref{fig:wave} are marked with asterisks in \Cref{fig:KHtest}. Notably, all three waves lie within the unstable region, supporting the feasibility of a KH-driven origin for the observed surface waves.

\section{Discussion and Conclusion}\label{sec:dis_conc}   % [~]

% - What are we doing in this work, and what do the results show? [-]
In this study, we examined the structure and dynamics of switchbacks (SB) boundaries in the young solar wind, with a particular focus on the presence of surface wave activity and its potential origin. Based on in-situ magnetic field and plasma measurements from Parker Solar Probe, we identified surface waves at SB boundaries and assessed their stability through the evaluation of the Kelvin-Helmholtz (KH) instability threshold. Our analysis indicates that, under certain conditions, the SB boundaries can support unstable configurations, consistent with KH-driven surface waves that may have developed locally at the boundary layer.

% - Interpretation of a circularly polarized wave in Figure 2 [~]
Although surface waves are often assumed to be linearly polarized, earlier theoretical work by \cite{Hollweg1982} reported that surface wave fluctuations can be elliptically, and occasionally circularly, polarized. One of the wave events identified in Figures \ref{fig:wave}(h) and \ref{fig:wave}(l) (07:48:43 UT and 1.75~Hz) is consistent with this expectation, showing a circularly polarized waveform with a wave vector nearly perpendicular to the boundary normal, suggesting surface-parallel propagation. The observed polarization is left-handed.
The circular polarization observed here may reflect the generation of MHD waves inside the structure, similar to what is observed at the flanks of the terrestrial magnetosphere during trains of KH-generated surface waves at the magnetopause \citep{agapitov2009}.  %\ke{mode conversion? boundary distortion? projection effect? -> need to check with Oleksiy}

\begin{figure*}
    \centering
    \includegraphics[width=\textwidth]{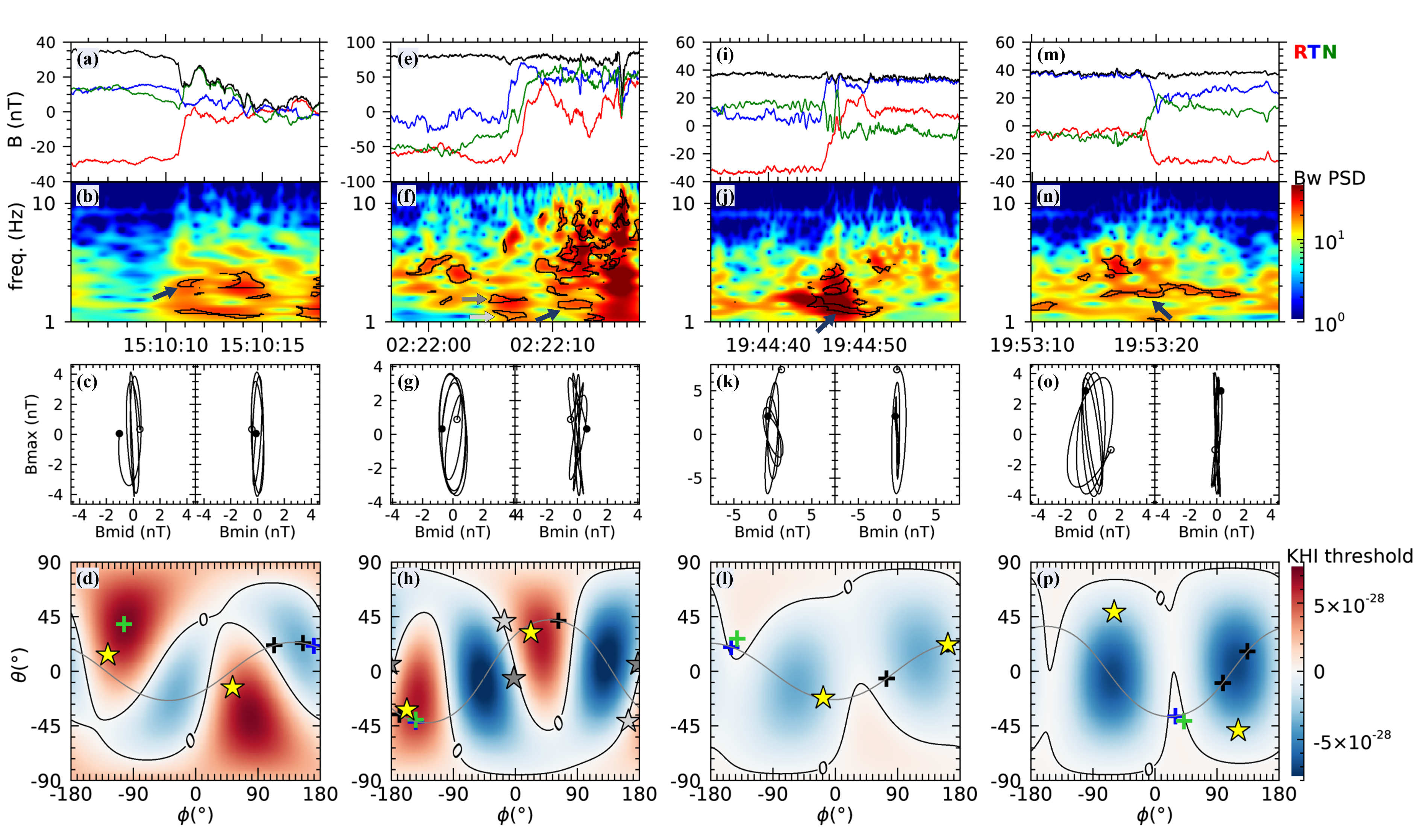}
    \caption{Summary of four additional surface wave events at SB boundaries. The first two events (panels a–h), observed at 15:10 UT on 2018 October 31 \citep{agapitov+2023} and 02:22 UT on 2018 November 8 \citep{farrell+2020}, are located in unstable regions of the KH threshold test, while the latter two events (panels i–p), observed at 19:44 UT and 19:53 UT on 2018 October 31 \citep{agapitov+2023}, are in stable regions. Each column corresponds to one event and consists of (from top to bottom): the magnetic field components, wavelet power spectral density (PSD), hodograph in the minimum variance plane, and the directional dependence of the KH instability threshold. In the PSD panels, contours indicate intervals that satisfy the linear surface wave,  $|\mathrm{ellipticity}| < 0.5$ and $\theta_{k\cdot n} > 60^\circ$.
    Hodographs correspond to selected waves at 2 Hz, 1.4 Hz (1.15 and 1.6 Hz), 1.2 Hz, and 1.75 Hz. Black arrows indicate the primary waves; light gray and gray arrows in the second event show additional components. Open and filled circles mark the hodograph's start and end points. The corresponding wave vectors are shown in the KH threshold panels using matching star symbols, consistent with \Cref{fig:KHtest}.}
    \label{fig:additional cases}
\end{figure*}

% - introduce for additional cases (Figure 4): Consistency across events [-]
To examine whether such features are consistently observed, we consider four additional events (as shown in \Cref{fig:additional cases}). All four cases, located at SB boundaries, show linearly or weakly elliptically polarized surface waves, as defined by $|\mathrm{ellipticity}| < 0.5$ and $\theta_{k\cdot n} > 60^\circ$. Among them, two events are found within unstable regions of the KH threshold diagram (Figures~\ref{fig:additional cases}(d) and \ref{fig:additional cases}(h)), while the other two are located in stable configurations (Figures~\ref{fig:additional cases}(l) and \ref{fig:additional cases}(p)).

In the second event (Figure~\ref{fig:additional cases}(e–h)), two additional wave components at 1.15 Hz and 1.6 Hz are identified at earlier times than the primary 1.4 Hz wave. Both are observed closer to the onset of the magnetic field rotation (\Cref{fig:additional cases}(e)), as also seen in the first, third, and fourth events of Figure~\ref{fig:additional cases}. Although the 1.4 Hz wave occurs slightly later, it remains within the transition layer approaching the SB. While the primary 1.4 Hz wave (yellow star in \Cref{fig:additional cases}(h)) is clearly located within the unstable region of the KH threshold diagram, the 1.15 Hz (the light grey star) and 1.6 Hz waves (the grey star) lie near the threshold boundary—just inside the unstable and stable regions, respectively.

% - Implication of surface waves: [-]
The presence of surface waves at the SB boundaries, which often exhibit characteristics of TDs, may have important implications for plasma dynamics both inside and around the SBs. In particular, if the boundary layer becomes unstable to the KH instability, it could enable particle exchange across the SB boundary. Such transport may occur even when the magnetic structure of the SB is nominally closed \citep{bizien+2023}, thereby allowing limited mixing between the internal and external plasma populations.

% - Implication of surface waves: the erosion of the boundaries [-] 
Moreover, the continued growth of KH instabilities can lead to the progressive erosion of SB boundaries, modifying the boundary morphology through gradual structural relaxation, which transforms initially sharp and simple boundaries into more broadened and structurally complex regions.
Observational indications of such erosion are presented by \citet{farrell+2021}, who report that boundaries with abrupt, step-like changes in magnetic and plasma parameters (e.g., \( B_r\), \( V_r\)) exhibit little wave activity, suggesting relatively young and unprocessed structures. 
In contrast, more degraded boundaries are accompanied by enhanced ultra-low frequency (ULF) fluctuations (0.07–1 Hz), which are lower in frequency than the surface waves examined in this study.
These findings from \citet{farrell+2021} imply that the sharpness of the SB boundaries can decrease over time through cumulative wave-induced erosion or instability-driven mixing. 
The results presented in this study are consistent with the interpretation, as the observed surface waves may represent remnants of KH-driven wave activity in the young solar wind. 
This activity may contribute to the structural evolution of the SB boundaries during their early-stage development.

%- comparisons with previous works: Similarities and key differences [-]
Previous studies have reported the presence of wave activity and coherent structures at SB boundaries \citep[e.g.,][]{krasnoselskikh+2020, mozer+2020, larosa+2021} without quantitatively evaluating the instability conditions.
This study provides observational support for the possibility that some surface waves arise from locally unstable shear layers. 
The enhanced shear within the SB, supported by strong perturbations in both the magnetic field ($\Delta \vec{B}$) and the plasma velocity ($\Delta \vec{V}$), may have further contributed to the onset of instability.

% - summarize our work [~]
In summary, we conducted a comprehensive investigation into the structure and stability of SB boundaries, with particular emphasis on the presence and origin of waves observed in the vicinity of the boundaries. Our key findings are summarized as follows:

\begin{enumerate}
\item The elevated wave activity observed in the frequency range of 1-5 Hz at SB boundaries exhibits characteristics typical of surface waves based on PSP observations at heliocentric distances of 35-55 \(R_S\). These waves propagate along the boundaries of SBs and along \(\Delta \vec{V}\), with their amplitudes rapidly decreasing as the distance from the boundary increases. The associated magnetic field perturbations display nearly linear polarization.

\item The boundaries of SBs may be susceptible to the generation of surface waves driven by the Kelvin-Helmholtz instability (KHI), suggesting that the observed wave activity at these boundaries results from shear flow instabilities.

%\item When \(\Delta \vec{B} = \pm \alpha \Delta \vec{V}\), and \(\alpha < 2\) (typically, for SBs, \(0.4 < \alpha < 1.1\)), the boundary remains stable. The observed bursts of wave activity are likely remnants of surface instabilities that were active closer to the Sun.
\item When \(\Delta \vec{B}\) and \(\Delta \vec{V}\) are nearly aligned, the boundary remains stable for the KHI, unless the velocity shear is more than twice the magnetic shear. However, in our observations on SBs' boundaries, the velocity shear typically ranges from 40\% to 90\% of the magnetic shear, indicating that the instability condition is not satisfied. The observed bursts of wave activity are thus likely remnants of surface instabilities that developed closer to the Sun.

\item The configuration of the magnetic field and velocity at the boundaries, which leads to instability, arises from deviations from the precise alignment of \(\Delta \vec{B}\) and \(\Delta \vec{V}\) in the nascent solar wind. The subsequent release of the KHI is hypothesized to facilitate the alignment of \(\Delta \vec{B}\) and \(\Delta \vec{V}\) at SB boundaries observed at heliocentric distances of 35-55 \(R_S\).
\end{enumerate}
% - Further analysis is required to... [-] 
Further work is needed to quantify the growth rate of the KH instability under varying solar wind conditions. Statistical analyses across different PSP encounters could also help assess the prevalence of KH-driven surface waves and their contribution to young solar wind.

\section{Acknowledgments}
KEC, LC, and OVA were supported by NASA contracts  80NSSC22K0433, 80NNSC19K0848, 80NSSC21K1770, and NASA’s Living with a Star (LWS) program (contract 80NSSC20K0218). 
NB and TD acknowledge support from CNES.
We thank the NASA Parker Solar Probe Mission, the SWEAP team led by Justin Kasper, and the FIELDS team led by Stuart Bale for the use of the data.
Parker Solar Probe was designed and built and is now operated by the Johns Hopkins Applied Physics Laboratory as part of NASA’s Living with a Star (LWS) program (contract NNN06AA01C)

\bibliography{ref}{}
\bibliographystyle{aasjournal}

\end{document}